\documentclass[journal=jacsat,manuscript=article]{achemso}

\usepackage[version=3]{mhchem} 
\usepackage[T1]{fontenc}       

\author{Tianhua Feng}
\affiliation{Department of Electronic Engineering, College of Information Science and Technology, Jinan University, Guangzhou, 510632, China}
\author{Wei Zhang}
\affiliation{Department of Electronic Engineering, College of Information Science and Technology, Jinan University, Guangzhou, 510632, China}
\author{Zixian Liang}
\affiliation{College of Electronic Science and Technology, Shenzhen University, Shenzhen, 518060, China}
\author{Yi Xu}
\affiliation{Department of Electronic Engineering, College of Information Science and Technology, Jinan University, Guangzhou, 510632, China}
\email{e_chui@qq.com}
\author{Andrey E. Miroshnichenko}
\affiliation{School of Engineering and Information Technology, University of New South Wales, Canberra, ACT, 2600, Australia}
\email{andrey.miroshnichenko@unsw.edu.au}

\title[An \textsf{achemso} demo]
  {Isotropic Magnetic Purcell Effect}

\keywords{Purcell effect, magnetic dipole, isotropic emission, all-dielectric nanocavity, magnetic dipole mode, dipole orientation}

\begin{document}


\begin{tocentry}

\includegraphics[scale=1]{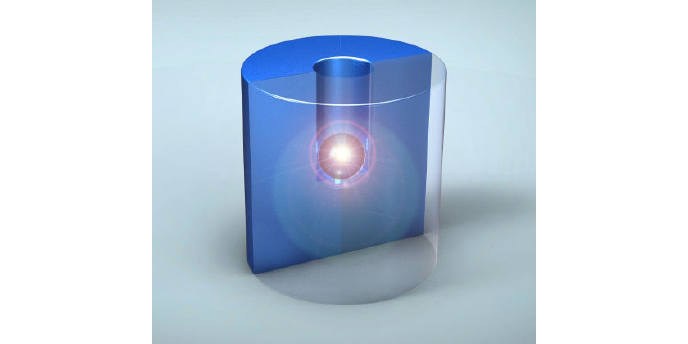}
\centering

\end{tocentry}

\begin{abstract}
Manipulating the spontaneous emission rate of optical emitters with all-dielectric nanoparticles benefits from their low-loss nature and thus provides relatively large extrinsic quantum yield. However, such Purcell effect greatly depends on the orientation of the dipole emitter. Here, we introduce the concept of \textit{isotropic magnetic Purcell effect} with Purcell factors about 300 and large extrinsic quantum yield (more than 80\%) for a magnetic dipole emitter of arbitrary orientation in an asymmetric silicon nanocavity. The extrinsic quantum yield can be even boosted up to nearly 100\% by utilizing a GaP nanocavity. Isotropy of the Purcell factor is manifested via the orientation-independent emission of the magnetic dipole source. This isotropic Purcell effect is robust against small displacement of emitter on the order of 10 nm, releasing the requirement of precise alignment in experiments.
\end{abstract}

\textbf{Keywords: }
Purcell effect, magnetic dipole, isotropic emission, all-dielectric nanocavity, magnetic dipole mode, dipole orientation


\clearpage
Magnetic Purcell effect (PE), which manifests itself as the manipulation of spontaneous decay rate of a quantum emitter with magnetic transitions, has received tremendous attention during these years\cite{LPR-Review,Zia_PRL,PRL_2014,PRL_2015,NC_Maser,APL_Belov}. In general, the interactions between natural materials and the magnetic component of light are very weak. Nevertheless, the magnetic component of light can play a crucial role in the realization of exotic functions \cite{Review_Meta,MagMirror,ACSNano_Isabelle,PRB_WLiu,OE_Liu,Optica_Kivshar,PRL2017}. Metamaterials have been used to enhance the magnetic response of materials in the optical spectrum, though most of them rely on the lossy metallic structures\cite{SRR_Giessen,decker}. Very recently, their all-dielectric counterparts were demonstrated to be a good alternative in nanophotonics because of their low-loss nature in the optical spectrum\cite{NT-Jacob,Sci-Andrey,NC2015,NL2017_GaP}. They can provide versatile platforms for engineering the magnetic light-matter interactions in the visible region\cite{MagLight,NL_MD,Optica2016,OL_2016,ACSPhoton2017}.

Magnetic dipole (MD) emission can be enhanced via metallic surface\cite{Zia_PRL}, split-ring resonator \cite{SRR_Giessen}, plasmonic nanoburger \cite{OL2011Feng} and diabolo nanoantenna \cite{ACSPhoton_diabolo}. However, considerable absorption of these metallic structures results in low extrinsic quantum yield in the visible spectrum. A better solution is to employ nano-resonators made from dielectric materials with high permittivity \cite{Sci-Andrey}. It has been proposed that a silicon (Si) nanosphere or structured nanodisk with magnetic resonance can effectively promote the MD emission \cite{DieAnt,PRB_Rolly,OL_2016,ACSPhoton2017}. In addition, Si nanodimers can further enhance this effect \cite{JPPC,NL1,NL2}. However, the Purcell factor (PF) is sensitive to the dipole orientation and location for all of the structures mentioned above\cite{SRR_Giessen,Vos_PRA,Wang_PRA}. This is because the MD emission enhancement relies on the excitation of a specific magnetic mode. In order to maximize the enhancement, the MD emitter should be located at the maximum of the magnetic hot-spot and its dipole moment should be parallel to the local magnetic field. There are some specific nanotechnologies, such as the host-guest chemistry, which can locate and align the light emitters on the single-molecule level\cite{Nature2016}. Unfortunately, for many widely-used nano-emitters, such as colloidal quantum dots and ions, controlling their dipole orientations pricisely is very challenging\cite{NP_2016}. Therefore, isotropic magnetic PE with dipole orientation-independent emission enhancement is in great demand, as it is important and practical for shaping the emission \cite{PRL_CWQiu,Ant_Alu}.

Here, we introduce a structured dielectric nanoparticle that can realize isotropic magnetic PE for a MD emitter in the visible spectrum. Such an asymmetric nanocavity supports two MD modes with their effective dipole moments being orthogonal to each other. By matching the PFs of the horizontally and vertically orientated MD emitters, we can realize isotropic magnetic PE.  

\section{RESULTS AND DISCUSSION}
\begin{figure}[!b]
 \includegraphics[width=0.8\columnwidth]{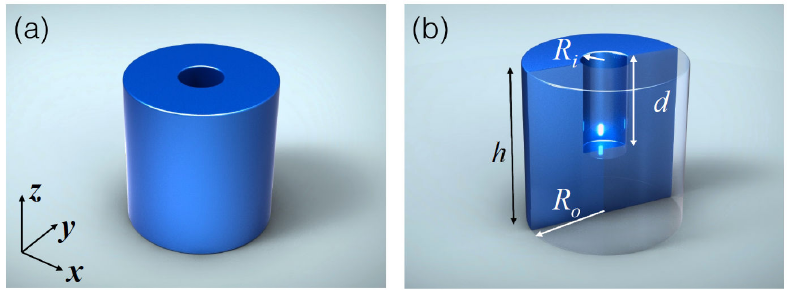}
 \centering
  \caption{(a) Schematics of the proposed silicon nanocavity (SNC). (b) Cross-section view of the SNC. The outer radius is $R_{o}$= 65 nm and the height is $h$ = 136 nm. The inner radius is $R_{i}$ = 19 nm and the depth of the void is $d$ = 78 nm.}
  \label{fgr:fig1}
\end{figure}

The proposed Si nanocavity (SNC) is schematically shown in Figure \ref{fgr:fig1}. A cylindrical air void is coaxially introduced in the upper region of a nanodisk. The cross section of the SNC is shown in Figure \ref{fgr:fig1}b by sketching part of the SNC with transparent materials for clarity. The geometric parameters can be found in the caption. It should be noted that similar structures have been employed to study the magnetoelectric coupling effect induced by geometric symmetry breaking\cite{Kivshar_PRB}. Here, we show that such a simple system continues to harvest functionality for nanophotonics.
In order to investigate the electromagnetic modes supported by the SNC, we conduct the full-wave simulations for plane wave excitation. It allows us to identify the electromagnetic modes of the SNC by employing the multipole decomposition approach \cite{NJP}:
\begin{equation}
\label{eq:eq1}
a_{lm}=\frac{(-i)^{l+1}kr}{h_l^{(1)}(kr)E_0[\pi(2l+1)l(l+1)]^{1/2}}\int Y^*_{lm} \frac{\textbf{r} \cdot \textbf{E}_s}{r}\textit{d}\Omega,
\end{equation}
\begin{equation}
\label{eq:eq2}
b_{lm}=\frac{(-i)^{l} Z kr}{h_l^{(1)}(kr)E_0[\pi(2l+1)l(l+1)]^{1/2}}\int Y^*_{lm} \frac{\textbf{r} \cdot \textbf{H}_s}{r}\textit{d}\Omega,
\end{equation}
where $a_{lm}$ and $b_{lm}$ are the electric and magnetic multipole scattering coefficients, $\textbf{E}_s$ and $\textbf{H}_s$ are the scattered fields, $r$ is the radius of a hypothetical sphere enclosing the SNC, $k$ is the wavenumber in the free space, $h_l^{(1)}(kr)$ is the first kind of Hankel functions, $Y^*_{lm}$ is the conjugated spherical harmonics, $E_0$ is the amplitude of the incident electric field, and $Z$ is the wave impedance of vacuum. In this way, we can obtain the scattering efficiency of each multipolar mode as:
\begin{equation}
\label{eq:eq3}
Q(l,m)=\frac{\pi}{k^2G} \sum_{l=1}^{\infty} \sum_{m=-l}^{l} (2l+1) [|a_{lm}|^2+|b_{lm}|^2],
\end{equation}
where $G$ is the corresponding geometric cross section along the specific direction. It should be noted that we sum the contributions from all \textit{m} for each multipole order \textit{l}. For the full-wave simulations, we employ the commercial FEM solver (COMSOL Multiphysics  5.2a), in which the Wave Optics Module is applied. The radius of the hypothetical sphere for multipole decomposition is $0.4\lambda_c$, where $\lambda_c=590$ nm. A spherical PML layer with thickness of $1/8\lambda_c$ and radius of $\lambda_c$ is used. The maximum mesh size is $1/10\lambda_c$ for vacuum and $1/35\lambda_c$ for the SNC, which are sufficient for obtaining the convergent results. 

\begin{figure}[!h]
 \includegraphics[width=0.8\columnwidth]{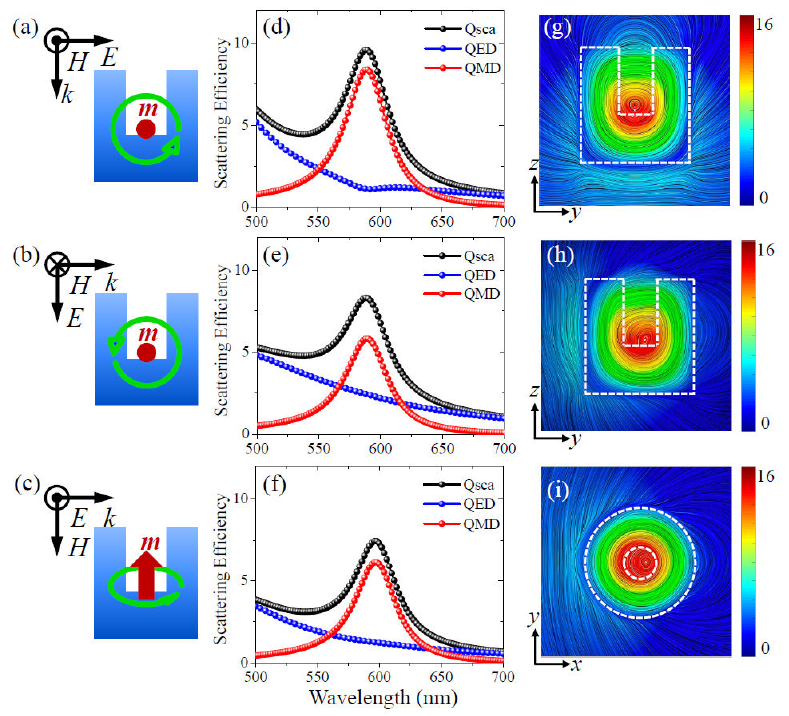}
  \caption{(a) - (c) The SNC excited by plane waves in three different scenarios. The MD modes can be excited with different dipole moment. The red arrows schematically show the MD moments while the green circular arrows show the displacement currents. (d) - (f) The total scattering efficiency (Qsca) and the contributions of the ED (QED) and MD (QMD) modes to the total scattering. (g) - (i) The distributions of magnetic field $|H|$ at the corresponding resonant wavelengths are presented with color map. The distribution of electric field is also shown with the streamlines superposed on the color map. The white dashed lines outline the profiles of the cross-section of the SNC.}
  \label{fgr:fig2}
\end{figure}

Considering the structural symmetry of the SNC, we choose three typical scenarios of plane wave excitations that are propagating along the $x$, $y$ and $z$ directions, as indicated in Figure \ref{fgr:fig2}a - c, respectively. Firstly, we consider the case when a plane wave is impinging on the SNC along the $-z$ direction with the $H_x$ polarization. As a result, a MD mode with its induced dipole moment along the $x$ direction will be excited. It can be found from the decomposition results in Figure  \ref{fgr:fig2}d that the resonant wavelength of the MD mode is at around 590 nm, which is also the wavelength of a MD transition of the Eu$ ^{3+}$ ions \cite{Zia_PRL}. We have also plotted the electric field pattern in Figure \ref{fgr:fig2}g, which manifests itself as circulating displacement currents associated with a magnetic hot-spot at the center. This optically-induced MD mode with its dipole moment along the $x$ direction can significantly enhance the emission of a MD emitter orientating along the same direction. This kind of MD mode can also be excited by a plane wave propagating along the $y$ direction with the $H_x$ polarization, as indicated in Figure \ref{fgr:fig2}b. We can also find a MD resonance at around 590 nm in Figure \ref{fgr:fig2}e. The electromagnetic field pattern is also provided in Figure \ref{fgr:fig2}h and it looks similar to the first case. For the third case, a plane wave with the $H_z$ polarization is impinging on the SNC along the $y$ direction as shown in Figure \ref{fgr:fig2}c. A MD resonance can also be found at 590 nm, as shown in Figure \ref{fgr:fig2}f. In this case, the induced magnetic dipole moment is along the $z$ direction, as validated in Figure \ref{fgr:fig2}i. Consequently, we can briefly classify the MD modes supported by the SNC as the horizontal and vertical ones. The horizontal one is mediated by the split ring-like cross section of the SNC while the vertical one is originated from the disk-like cross section of the SNC \cite{OL_2016}, which are schematically shown in Figure \ref{fgr:fig2}a - c. Despite of the asymmetric geometry of the SNC, the horizontal and vertical MD modes can still spectrally overlap, as shown in the middle column of Figure \ref{fgr:fig2}. It is these modes that provide the opportunity to enhance the emission simultaneously for both horizontally and vertically orientating MD emitters, making it possible to realize the dipole orientation-insensitive emission enhancement. 

\begin{figure}[!t]
\centering
 \includegraphics[width=0.8\columnwidth]{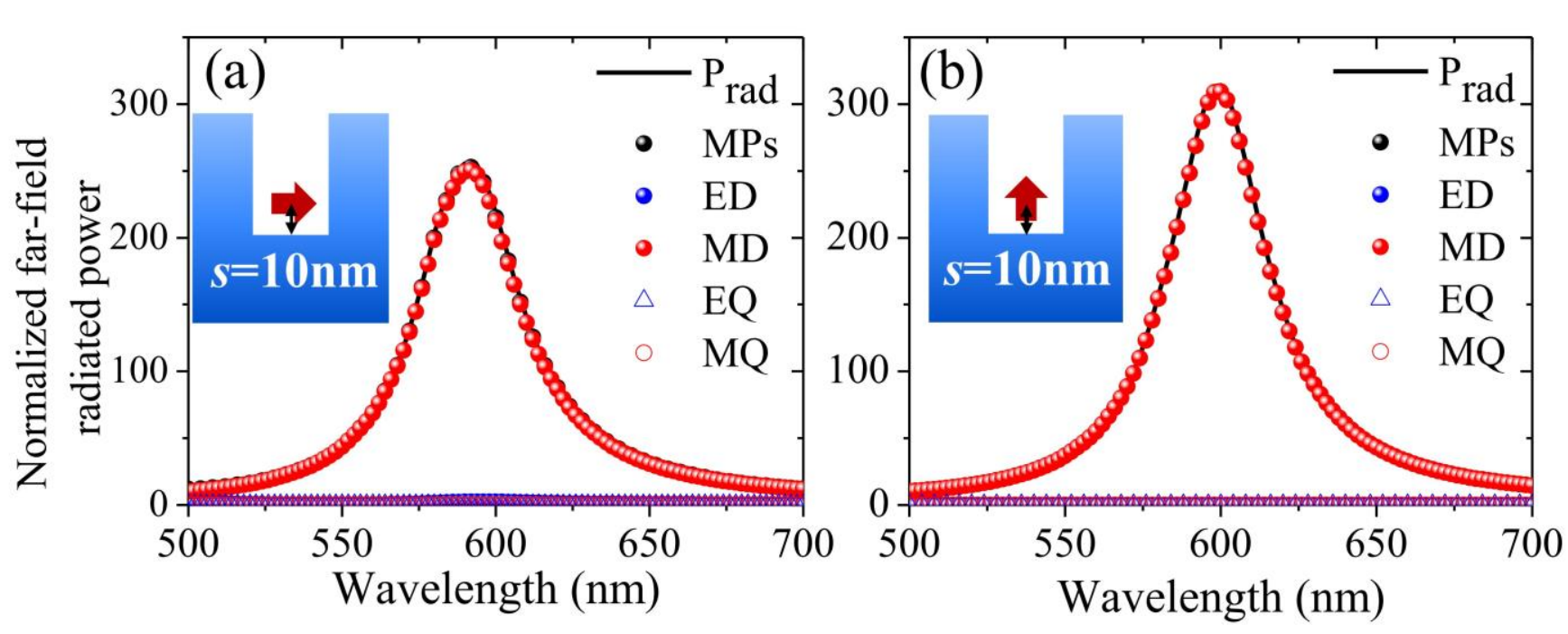}
  \caption{(a) and (b) Normalized far-field radiated power ($P_{rad}$) and that of each multipole mode (EQ: electric quadrupole; MQ: magnetic quadrupole) for two kinds of MD orientations, respectively. MPs: the sum of normalized radiated powers from the four multipole modes. Schematics of a MD emitter (red arrow) located inside the SNC with either horizontal or vertical dipole orientation are shown in the insets, respectively. In both cases, the emitter is located above the bottom of the void with a distance of $s$ = 10 nm. }
  \label{fgr:fig3}
\end{figure}

It should be noted that the SNC also supports electric dipole (ED) mode acompanying with the MD mode under the plane wave excitation, although its contribution to the total scattering is smaller than the MD one \cite{PRL2017}. Therefore, it is necessary to clarify how these modes are excited by a MD source other than a plane wave. In the following, we examine the radiated electromagnetic fields from a MD emitter in the SNC within the framework of multipole expansion\cite{Jackson}
\begin{equation}
\label{eq:eq4}
a_E(l,m)=-\frac{k}{Z h_l^{(1)}(kr)\sqrt{l(l+1)}}\int Y^*_{lm} \textbf{r}\cdot \textbf{E}_{tot} \textit{d}\Omega,
\end{equation}
\begin{equation}
\label{eq:eq5}
a_M(l,m)=\frac{k}{h_l^{(1)}(kr)\sqrt{l(l+1)}}\int Y^*_{lm} \textbf{r}\cdot \textbf{H}_{tot} \textit{d}\Omega,
\end{equation}
where $a_E$ and $a_M$ are the electric and magnetic multipole coefficients, $\textbf{E}_{tot}$ and $\textbf{H}_{tot}$ are the total radiated electromagnetic fields outside the SNC. Other parameters are similar with that in Eq. (\ref{eq:eq1}) and (\ref{eq:eq2}). As a result, we can calculate the coupling between the MD emitter and each multipole mode by evaluating their corresponding far-field radiated powers as
\begin{equation}
\label{eq:eq6}
P_{E(M)}(l,m)=\frac{Z}{2k^2}|a_{E(M)}(l,m)|^2.
\end{equation}
The power is normalized to that of a MD emitter without any structure in free space for the ease of comparison. As typical scenarios, we investigate two kinds of dipole orientations for the MD emitter, which are schematically shown in the inserts of Figure \ref{fgr:fig3}a and b. We present the normalized far-field radiated power of each multipole mode for both horizontal and vertical dipoles in Figure \ref{fgr:fig3}a and b, respectively. One can clearly see that the far-field radiated power from the MD emitter is almost coupled to the MD mode for both cases, whose resonant wavelengths are similar to the cases of plane wave excitations in Figure \ref{fgr:fig2}. The radiative contributions via other multipole modes are negligible. In addition, the result of the normalized far-field radiated power ($P_{rad}$), which is directly obtained from the integral of power flow through a closed surface containing the dipole and the SNC, nearly overlaps with the normalized radiated power of the MD mode. By summing the normalized radiated power of each multipole modes up to the quadrupole one (MPs) and comparing it with the directly calculated power $P_{rad}$, we can also benchmark our multipole expansion method and demonstrate the dominating MD mode radiation. We also further confirm our conclusion via the decomposition results based on the induced currents of the SNC (see Supporting Information).

\begin{figure}[!t]
 \includegraphics[width=0.8\columnwidth]{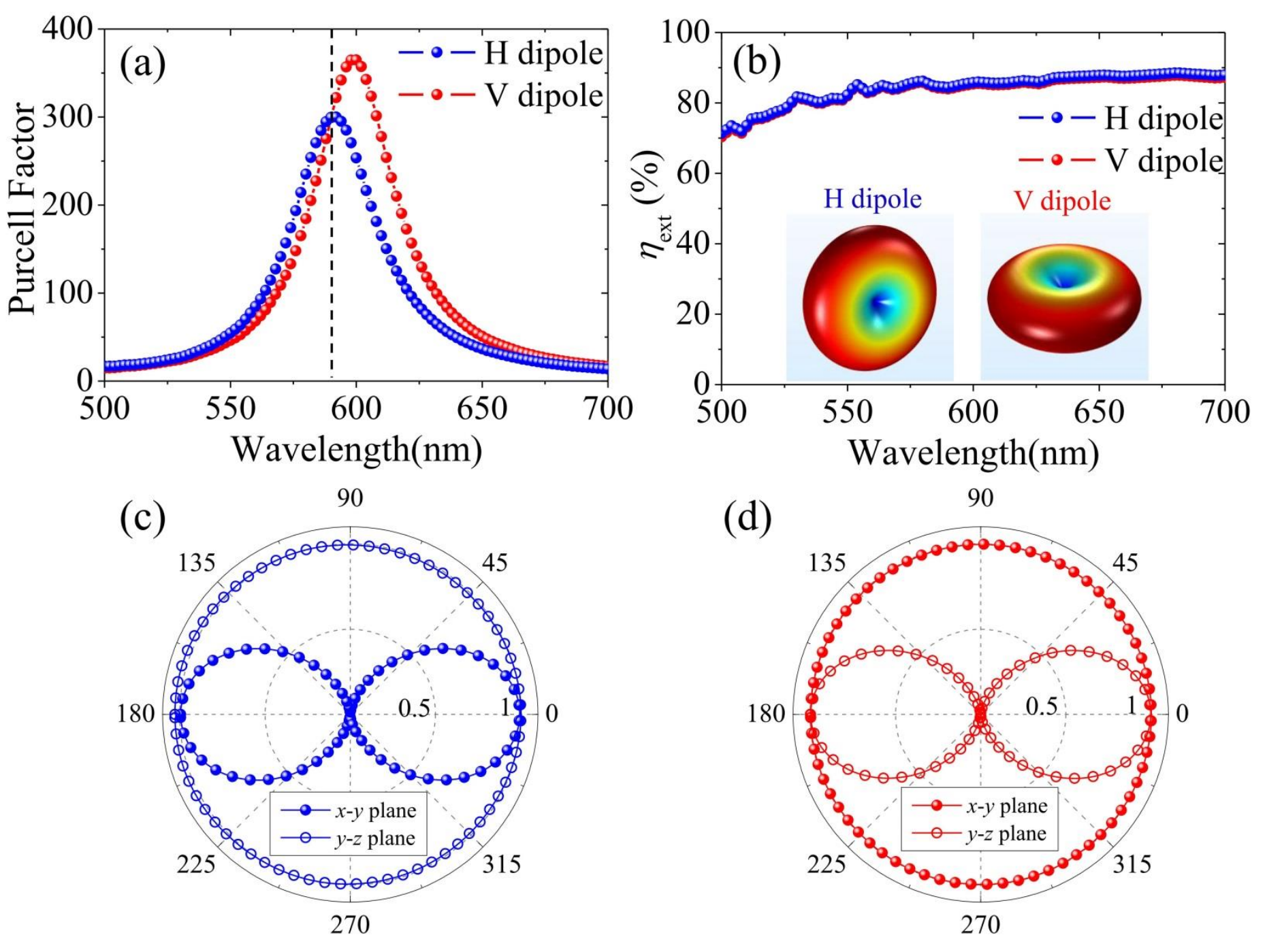}
  \caption{(a) PF spectra of the horizontal and vertical MD emitters. (b) The corresponding extrinsic quantum yield. The inserts show the 3D far-field radiation patterns of horizontal (left) and vertical (right) MD respectively at 590 nm. (c) and (d) 2D cross-section views of the far-field radiation patterns on the $x-y$ and $y-z$ planes for both cases.}
  \label{fgr:fig4}
\end{figure}

As the next step, we calculate the PF of the MD emitter by evaluating the ratio of the power lost by an ideal oscillating MD with and without the SNC, i.e. $PF = P_{w}/P_{w/o}$, where $P_{w}$ includes the far-field radiated power $P_{rad}$ and the one absorbed by the SNC $P_{abs}$. In order to characterize how the far-field radiated power would change once the lossy SNC was introduced, we study the extrinsic quantum yield which is defined as $\eta_{ext}= P_{rad}/(P_{rad}+P_{abs})$ with the assumption of unit intrinsic quantum yield for the sake of simplicity\cite{PRL2006}. The calculation of the far-field radiated power $P_{rad}$ is mentioned above while the absorbed power $P_{abs}$ is evaluated by the volume integral of power dissipation in the SNC. We show in Figure \ref{fgr:fig4}a the PF as a function of wavelength for both kinds of dipole orientations. The SNC is the same as the one shown in Figure \ref{fgr:fig3}. It shows that the PFs, which can be as high as 299, of the horizontally and vertically orientating MD emitters are the same at the wavelength of 590 nm although their resonant wavelengths are slightly different. It means that we can realize the \textit{isotropic magnetic PE} at 590 nm. We also study the extrinsic quantum yield of the MD emitter, as shown in Figure \ref{fgr:fig4}b. Because of the low-loss nature of Si material, the value of $\eta_{ext}$ can reach about 85\% for both orientations. As mentioned above, the emitted power is totally coupled to the MD modes of the SNC. This can be verified again by the doughnut-like 3D far-field radiation patterns and their corresponding 2D cross-section views, as shown in Figure \ref{fgr:fig4}b - d.

\begin{figure}[!t]
\centering
 \includegraphics[width=0.8\columnwidth]{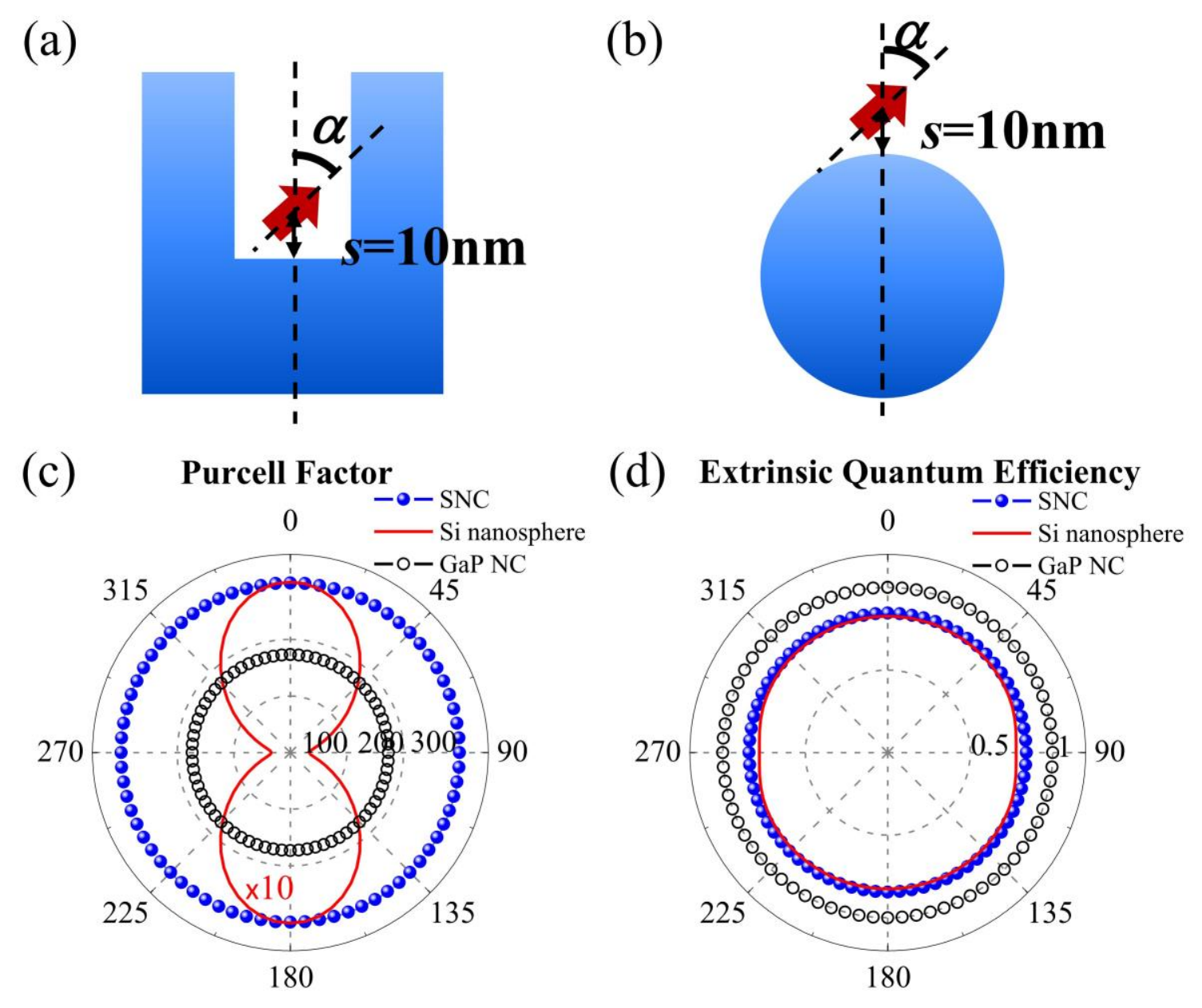}
  \caption{(a) Schematics of a MD (red arrows) located inside the SNC with a distance of 10 nm above the inner bottom. The dipole orientation is specified with a tilt angle $\alpha$ with respect to $z$ axis (dashed line). (b) The similar configuration for a Si nanosphere. (c) and (d) The PF and $\eta_{ext}$ as functions of the tilt angle $\alpha$ at the wavelength of 590 nm for the SNC (blue), GaP NC(black) and Si nanoparticle (red) cases. The parameters of the GaP NC are $R_{o}$ = 77 nm, $h$ = 150 nm, $R_{i}$ = 19 nm, $d$ = 85 nm and $s$ = 10 nm. The PF for the case of sphere is enlarged with 10 times for the ease of comparison.}
  \label{fgr:fig5}
\end{figure}

We further investigate how the PF changes when varying the orientation of the MD emitters. Considering the cylindrical symmetry of the SNC referred to the $z$ axis, we only need to study the case of a MD emitter with a tilt angle ($\alpha$) with respected to the $z$ direction, as shown in Figure \ref{fgr:fig5}a. The PFs as functions of $\alpha$ are shown in Figure \ref{fgr:fig5}c. It can be seen that the PFs are the same for all dipole orientations with a value of 299, indicating the realization of \textit{isotropic magnetic PE}. In Figure \ref{fgr:fig5}d, we also show the corresponding $\eta_{ext}$. It can be found that the extrinsic quantum yield for every angle is nearly the same too, which is as high as 85\%, indicating the isotropic magnetic PE with isotropic extrinsic quantum yield. For comparison, we also provide the results of the MD emission enhanced by a typical Si nanosphere with a radius of 70 nm, which is sketched in Figure \ref{fgr:fig5}b. We can see from Figure \ref{fgr:fig5}c that the PFs are anisotropic and show strong dependence on the dipole orientation, although the corresponding extrinsic quantum yield shown in Figure \ref{fgr:fig5}d is quasi-isotropic. Meanwhile, the maximum PF is only about 30, which is just ten percent of the SNC case. If we used GaP, whose loss is smaller than Si in the visible spectrum, to construct the nanocavity, the extrinsic quantum yield could be boosted up to nearly 100\% though the PF will be degraded, as shown in Figure \ref{fgr:fig5}c and d. It can be attributed to the lower refractive index of GaP, which will decrease the quality factors of the MD modes. The magnetic field enhancement mediated by the MD modes is also decreased compared to the case of the silicon nanocavity (see Supporting Information). In practical experiments, the emission enhancement would be greatly degraded if it was sensitive to the position of the emitter. For our proposed SNC, it is found that the dipole orientation-resolved PF just slightly changes even when the emitter is much closed to the inner wall of the void ($\sim$ 2 nm) with a distance-to-bottom of $s=$ 10 nm. Furthermore, we have also examined the cases of shifting the emitter up and down ($\sim$ 5 nm) and the isotropic magnetic PE is robust. It can be attributed to the quasi-homogenous distribution of magnetic field of the MD modes. Therefore, such results demonstrate that the isotropic magnetic PE is quite robust against the position deviation of the MD emitter within such ranges. Furthermore, the magnetic dipole emitters should be embedded in a specific kind of host material, such as Y$ _2$O$ _3$, in practical implementations\cite{Ref_Y2O3}. We have studied this situation and found that the isotropic magnetic PF can still be achieved (see  Supporting Information). We have also found that for the same SNC, the PF of an ED emitter is two orders of magnitude  smaller than that of a MD emitter, which can facilitate the differentiation of the MD and ED emission in experiment.

\section{CONCLUSION}
In summary, we have numerically demonstrated that the isotropic magnetic Purcel effect with isotropic extrinsic quantum yield can be realized with an asymmetric all-dielectric nanocavity. We reveal that the equal coupling between the MD emitter and the horizontal and vertical MD modes of the nanocavity is the crucial point to realize this isotropic emission. In addition, we find that such isotropic emission is quite robust against the location of the MD emitter on the order of ten nanometers. Our results might facilitate the manipulation of magnetic dipole emission utilizing all-dielectric platforms.

\begin{suppinfo}
Decomposition results based on the induced current; dependence of PF spectra on dipole orientation; role of imperfections; magnetic field distribution of the GaP nanocavity excited by a plane wave; PF of a MD emitter embedded in Y$ _2$O$ _3$ nanocrystals; PF of an ED emitter with SNC; validation of calculation methods.
\end{suppinfo}

\begin{acknowledgement}

T. F and Y. X acknowledge T. Guo for useful discussion. Financial support by the National Natural Science Foundation of China (NSF) (11704156, 11674130 and 61505114), Natural Science Foundation of Guangdong Province, China (2016A030308010, 2016TQ03X981, 2014A030313376 and 2014KQNCX128) and the Fundamental Research Funds for the Central Universities (21617346) are gratefully acknowledged. T. F. and W. Z. also thank the financial support of the leading talents of Guangdong province Program (00201502). The work of A. E. M was supported by the Australian Research Council. 

\end{acknowledgement}



\end{document}